\documentclass[letterpaper,11pt]{article}
\usepackage{jheppub}
\usepackage{physics}
\usepackage{relsize}
\usepackage{tikz}
\usepackage{pgfplots}
\usepackage{comment}

\usepackage[utf8]{inputenc}

\title{$T\bar{T}$ deformed YM$_{2}$ on general backgrounds from an integral transformation}
\author[a]{Aurora Ireland}
\author[b]{Vasudev Shyam}
\affiliation[a]{University of Chicago, Department of Physics,\\ 5640 S. Ellis Ave., Chicago, IL, 60637, USA}
\affiliation[b]{Perimeter Institute for Theoretical Physics,\\ 31 N. Caroline St., Waterloo, ON, N2L 2Y5, Canada}
\date{\today}

\begin{document}

\abstract{
We consider the $T\bar{T}$ deformation of two dimensional Yang--Mills theory on general curved backgrounds. We compute the deformed partition function through an integral transformation over frame fields weighted by a Gaussian kernel. We show that this partition function satisfies a flow equation which has been derived previously in the literature, which now holds on general backgrounds. We connect ambiguities associated to first derivative terms in the flow equation to the normalization of the functional integral over frame fields.
We then compute the entanglement entropy for a general state in the theory. The connection to the string theoretic description of the theory is also investigated.}
\maketitle

\section{Introduction}
The $T\bar{T}$ deformation, introduced in \cite{Smirnov:2016lqw},\cite{Cavaglia:2016oda}, is a solvable irrelevant operator deformation of quantum field theories in two dimensions. The deforming operator is defined by the coincidence limit of the following bi-local operator formed from the energy momentum tensor:
\begin{equation}
T\bar{T}(x)\equiv \lim_{y\rightarrow x}\epsilon^{\alpha\beta} \epsilon^{\mu\nu} T_{\mu\alpha}(x)T_{\nu\beta}(y).
\end{equation}
It was shown in \cite{Zamolodchikov:2004ce} that this coincidence limit defines a local operator up to total derivatives. Upon deforming a quantum field theory by such an operator, the result is a one-parameter family of quantum field theories whose partition functions $Z$ solve the following flow equation:
\begin{equation}
\partial_{\mu}Z=\int \textrm{d}^{2}x \sqrt{g} \langle T\bar{T}(x)\rangle Z,\label{int_flow_eqn}
\end{equation}
provided the initial condition 
\begin{equation}
Z|_{\mu=0}=Z_{0},
\end{equation}
where $Z_{0}$ is the partition function of the undeformed theory.

With this definition of the deformation, we can in principle discuss the $T\bar{T}$ flow of theories living on general curved backgrounds, for which the expectation value of the deforming operator is given by the following expression:
\begin{equation}
\langle T\bar{T}(x)\rangle= \lim_{x\rightarrow y} \frac{1}{Z[g]}\frac{\epsilon^{\alpha\beta}\epsilon^{\mu\nu}}{\sqrt{g}(x)\sqrt{g}(y)}\frac{\delta^{2}Z[g]}{\delta g^{\mu\alpha}(x) \delta g^{\nu\beta}(y)}.\label{operator_defn}
\end{equation}
In practice, it isn't guaranteed that the quantity above is computable, so one must seek specific contexts wherein it is. 
In particular, when the state of interest lacks translation invariance, one cannot exploit the factorization argument of \cite{Zamolodchikov:2004ce} to establish the unambiguous definition of this operator. There could potentially be curvature dependent terms in the OPE that make it ambiguous as to what the local operator obtained as a result of the coincidence limit of the bilocal operator above is. Many of these difficulties can be evaded in the classical limit of the theory under consideration, and this was the method used in \cite{Bonelli:2018kik} to obtain the deformation of various classical Lagrangians.

When the undeformed theory is a conformal field theory (CFT), the $T\bar{T}$ flow coincides with a renormalization group flow. Given that the operator of interest is irrelevant, it triggers a flow from the CFT in the infrared (IR) to some other theory in the ultraviolet (UV). The RG flow essentially runs backwards from that deformed theory to the CFT along the UV critical surface. 

In \cite{McGough:2016lol}, it was argued that the $T\bar{T}$ deformation of holographic conformal field theories are the field theory dual to gravity in AdS$_{3}$ with a finite radial cutoff surface. This proposal has undergone many checks (\cite{Kraus:2018xrn}, \cite{bbc2019}, \cite{Donnelly:2018bef}, and references therein). There have also been higher dimensional generalizations, which were proposed and studied in \cite{Taylor:2018xcy},\cite{Hartman:2018tkw}, and \cite{Shyam:2018sro}. Additionally, the authors of \cite{Gross:2019ach} have studied the generalization of this deformation in quantum mechanics. The supersymmetric generalization of such an operator has been studied in \cite{Baggio:2018rpv}, \cite{Chang2019}, \cite{Chang:2019kiu}, \cite{Coleman:2019dvf}, \cite{Cribiori:2019xzp} and \cite{Jiang:2019hux}. 

From another point of view, one can define the theory on a cutoff surface of AdS$_{3}$ as one whose partition functions solves the bulk radial Wheeler-de Witt (WdW) equation. Based on this intuition, an integral transformation was found in \cite{Freidel:2008sh} that relates the CFT partition function with a solution to the radial WdW equation. This relation between the $T\bar{T}$ deformation and solutions to the WdW equation was also speculated in \cite{Caputa:2019pam}. The prescription is to integrate over the dyads of the space on which the undeformed theory lives with a certain Gaussian kernel in order to obtain this wavefunction. Explicitly it reads:
\begin{equation}\label{deformation}
    Z[f] = \int \mathcal{D}e \,  \exp \left[ - \frac{1}{\mu} \int \textrm{d}^{2}x \epsilon^{\alpha \beta} \epsilon_{ab} (e - f)^a_\alpha (e - f)^b_\beta \right] \, Z_0[e] \,.
\end{equation}
Here, $f^{a}_{\mu}$ is the dyad, or frame field, associated to the geometry on which the $T\bar{T}$ deformed theory lives, while $e^{a}_{\mu}$ is the dyad on the geometry the undeformed theory inhabits. Note that these are functional integrals over dyads on spaces with fixed topology. 

Defining the partition function of the deformed theory in this way ensures that the equation \eqref{int_flow_eqn} with the $T\bar{T}$ operator defined as \eqref{operator_defn} is satisfied as a functional identity. In other words, the equation 
\begin{equation}
\partial_{\mu}Z = \int_{x}\textrm{d}^{2}x \epsilon^{ab}\epsilon_{\mu\nu} :\frac{\delta^{2}Z}{\delta f^{a}_{\mu}(x)\delta f^{b}_{\nu}(x)}:,\label{fo_fe}
\end{equation}
which is a rewriting of the flow equation in first order variables, follows directly as an identity from \eqref{deformation}. \footnote{Note that in the first order variables, we define a mixed index stress tensor 
\begin{equation}
T^{\mu}_{a}= T^{\mu\nu} f_{\nu a},
\end{equation}
in terms of which
\begin{equation}
T\bar{T}(x) = \frac{1}{2}\epsilon_{\mu\nu}\epsilon^{ab} T^{\mu}_{a}T^{\nu}_{b}(x) = \textrm{det}T(x)
\end{equation}}
More specifically, the coincidence limit in the definition \eqref{operator_defn} can be taken provided a normal ordering prescription is followed when taking the functional derivatives. This is what the semicolons in the expression \eqref{fo_fe} denote:
\begin{equation}
\epsilon^{ab}\epsilon_{\mu\nu} :\frac{\delta^{2}}{\delta f^{a}_{\mu}(x)\delta f^{b}_{\nu}(x)}: = \epsilon^{ab}\epsilon_{\mu\nu} \frac{\delta^{2}}{\delta f^{a}_{\mu}(x)\delta f^{b}_{\nu}(x)}+ \frac{2}{\lambda}\delta^{(2)}(0). 
\end{equation}

For details of this calculation, we refer the reader to section 2.2. of \cite{mazenc2019t}. 
All this, in turn, means that the problem of having to carefully define the operator \eqref{operator_defn} is traded for that of having to define the functional integral \eqref{deformation} over dyads. This means that the above functional integral definition holds even when the undeformed theory doesn't possess conformal symmetry. Furthermore, we will see that ambiguities associated to operator ordering find an analogous manifestation in the choice of the normalization factor of the frame-field path integral measure in \eqref{deformation}.
                                    
The connection between this transformation and the $T\bar{T}$ deformation was proposed in \cite{McGough:2016lol}. This perspective was further elucidated in \cite{tolley2019t}, where it was noted that the action appearing in the exponent of \eqref{deformation} is that of ghost free massive gravity in two dimensions. The aforementioned fact that \eqref{deformation} can be used to define the deformed partition function of theories without conformal symmetry is a reflection of the fact that massive gravity can be coupled to a quantum field theories without conformal symmetry. In particular we will apply it to two dimensional Yang--Mills with gauge group $U(N)$, whose partition function has particularly simple dependence on the background geometry. In doing so, we shall obtain a deformed partition function on an arbitrary curved background, where the Hamiltonian we infer of the deformed theory is identical to the one obtained from the classical analysis of \cite{Conti2018}.
On the other hand, we will see what subtleties accompany the use of this method of defining $T\bar{T}$ deformed theories.

We should mention that this is indeed similar to the prescription of \cite{Dubovsky:2017cnj},\cite{Dubovsky:2018bmo}, where it was argued that the $T\bar{T}$ deformation arises from coupling the undeformed theory to a particular dilaton-gravity theory in two dimensions known as Jackiw--Teitelboim gravity. This perspective allowed the authors of these articles to derive the CDD phases, which encode the deformation of the S-Matrix, as well as the torus partition function. The precise connection between these approaches is covered in \cite{mazenc2019t}.

\section*{Comparison to earlier work}
The deformed Lagrangian and Hamiltonian densities for two dimensional Yang--Mills theory were first presented in \cite{Conti2018}. These authors argued that the effect of the $T\bar{T}$ deformation could then be incorporated via a simple redefinition of the quadratic Casimir eigenvalues. They also constructed a flow equation for the deformed partition function written in terms of area derivatives, which matches with that of \cite{Cardy2018} for the case of the torus. \cite{Santilli2019} expanded upon this work by studying the deformed partition function for the case of YM$_2$ on the two-sphere. By performing the large $N$ analysis, they determined that the Douglas--Kazakov phase transition induced by unstable instantons persists in the deformed theory for a range of deformation parameter values.

Here, in section \ref{ym}, we present another means of deriving the deformed partition function which matches with the result of \cite{Conti2018} and \cite{Santilli2019}. Thanks to the Gaussian nature of the integral in Eq.\,(\ref{deformation}), it can be performed exactly to obtain a result valid on a general background, provided an appropriate choice of measure for the path integration. This allows us to generalise the above results to arbitrary, possibly curved manifolds. Another advantage of our approach is that it provides a way to obtain the deformed Hamiltonian, previously obtained via classical analysis, directly in the quantum theory. 

We then derive a flow equation satisfied by the partition function on a general background, which matches that obtained in \cite{Cardy2018} when specialised to the case of the torus. It turns out that the integral kernel definition offers an interesting perspective on the presence of contact terms appearing in the generalised flow equation, which we comment on. Specifically, we make concrete the connection between operator ordering ambiguities in the flow equation and the choice of normalisation in the integral transform method.

To be clear, the assumption made here is that the deformation of the Hamiltonian that the classical analysis of \cite{Conti2018} provides indeed retains its form even in the quantum theory on a general background. At this point, there aren't any complementary methods to check whether this is the case other than on the torus or the cylinder.

Then, in section \ref{e&s}, we use our result to explore further YM$_2$ phenomena and determine how they are altered in the $T\bar{T}$ deformed theory. In particular, we look at the effect of the deformation on the string theoretic interpretation of YM$_2$, an analysis of which is currently lacking in the literature. As argued by \cite{Gross:1993}, YM$_2$ admits an effective string description in the large $N$ regime. We determine to what extent this description remains valid in the deformed theory. Finally, we use our main result of the deformed partition function on general background to compute entanglement entropy for an arbitrary state. We then specialise to the case of the Hartle--Hawking vacuum state in order to compare with entanglement entropy calculations for the undeformed theory in the existing literature \cite{Donnelly:2014}, \cite{GROMOV201460}.

\section{2D Yang Mills}\label{ym}

In order to demonstrate the applicability of the integral kernel definition of $T\bar{T}$ to general quantum field theories, we consider as a test case 2-dimensional Yang--Mills theory (YM$_2$). There are a number of reasons as to why YM$_2$ is ideal for these purposes. On one hand, it's tractable. The theory is semi-topological, depending only on the total area of the background manifold and the Euler characteristic characterizing its topology. This allows it to be solved exactly by topological field theory methods. Another consequence is that it has no local degrees of freedom. This renders it UV finite, so divergences are not an issue. Nevertheless, it has a number of interesting non-trivial features. For one, YM$_2$ admits a string theory interpretation in the large $N$ limit \cite{Gross:1993}. When the background manifold is a sphere, it also exhibits a (third order) phase transition induced by unstable instantons \cite{Douglas:1993}. Most importantly for our purposes, the partition function of YM$_2$ in representation basis has a very simple exponential dependence on the area. When written in the Vielbein formalism, it is quadratic in these frame fields. As such, the integral in Eq.\,(\ref{deformation}) is Gaussian and may be evaluated exactly, provided the appropriate measure for the path integration over zweibeins is chosen.

\subsection{$T\bar{T}$ deforming $Z_{YM}$}

For Yang--Mills theory living on a 2-dimensional manifold $\mathcal{M}$ of Euler characteristic $\chi$, the partition function $Z_{YM}$ admits the following group theory expansion \cite{Cordes:1994}:
\begin{equation}\label{Z0}
    Z_{YM} = \sum_\mathcal{R} (\text{dim} \, \mathcal{R})^\chi \, e^{- \frac{\lambda C_2}{2 N} A}\,.
\end{equation}
Here, the sum runs over all equivalence classes of irreducible representations $\mathcal{R}$ of the gauge group, which we will take to be $U(N)$. $C_2(\mathcal{R})$ is the quadratic Casimir eigenvalue associated with $\mathcal{R}$, $\lambda$ is the dimensionful 't Hooft coupling $\lambda = g_{YM}^2 N$, and $A$ is the total area of $\mathcal{M}$. To make contact with the form of the kernel, which is expressed in terms of zweibeins, we will write the area as an integral over the 2D volume 2-form, $A = \int \textrm{d}^{2}x f$, where $f \equiv \det f^a_\alpha(x)$. More explicitly, we have $\det f^a_\alpha(x) = f_\alpha^+ f_\beta^- - f_\beta^+ f_\alpha^- \equiv f^+ \wedge f^-$, where $f^\pm$ are the usual $f^\pm = (f^0 \pm f^1)/\sqrt{2}$.

Because it is quadratic in the frame fields, the YM$_2$ partition function presents a scenario for which the Gaussian integral involved in the kernel definition can be computed exactly. Of course, this relies on choosing a measure on the space of zweibeins that respects both the linearity property, i.e. 
\begin{equation}
\mathcal{D}e=\mathcal{D}(e+f) \,,
\end{equation}
and diffeomorphism invariance. As explained in Appendix A of \cite{tolley2019t}, such a measure can indeed be found, and it is defined with respect to the following supermetric on the space of zweibeins:
\begin{equation}
\delta s^{2}= -\int \textrm{d}^{2}x\, \epsilon^{\mu\nu} \epsilon_{ab} \delta e^{a}_{\mu}(x) \delta e^{b}_{\nu}(x)= - 2 \int \textrm{d}^{2}x\, \det[\delta e^{a}_{\mu}(x)] \,.
\end{equation}

Having chosen this measure, from eq.\,(\ref{deformation}), we can then obtain the $T\bar{T}$ deformed partition function $Z$:
\begin{equation}
    \begin{split}
        Z[f] & = \int \mathcal{D}e \, K[e,f] \, Z_0[e]\\
        & = \sum_\mathcal{R} (\text{dim} \, \mathcal{R})^\chi \int \mathcal{D}e \, e^{- \frac{1}{\mu} \int\textrm{d}^{2}x\, (e-f)^+ \wedge (e-f)^-} e^{- \frac{\lambda C_2}{2 N} \int \textrm{d}^{2}x\, e^+ \wedge e^-}\\
        & =  \sum_\mathcal{R} (\text{dim} \, \mathcal{R})^\chi \mathcal{N}_{\mathcal{R}} \sqrt{\frac{\mu \pi}{1+ \mu \lambda C_2/2N}} \, e^{- \frac{\lambda}{2N} \left( \frac{C_2}{1+ \mu \lambda C_2/2N} \right) \int \textrm{d}^{2}x\, f^+ \wedge f^-} \,,
    \end{split}
\end{equation}
where $\mathcal{N}_{\mathcal{R}}$ is a normalisation constant coming from our agnosticism regarding the functional measure. Note that since we are doing one Gaussian integral per representation, this constant could possibly depend on the quadratic Casimir $C_{2}(\mathcal{R})$, in addition to depending on $\mu$. We will see that it is important to be able to fix the normalization separately for each Gaussian integral.

In agreement with the claims of \cite{Conti2018}, we see that the quadratic Casimir eigenvalues are indeed dressed by the deformation as: 
\begin{equation}
    C_2 \rightarrow \frac{C_2}{1 + \mu \lambda C_2 / 2 N} \,.
\end{equation}

In order to normalise $Z[f]$, we look at the topological limit. In a topological theory, the energy-momentum tensor should vanish, such that the action of the $T\bar{T}$ deformation is trivial. We will thus fix $\mathcal{N}_\mathcal{R}$ by requiring that the integral transform of the topological theory yields the undeformed topological theory, $Z_0[A = 0] \overset{T\bar{T}}{\longrightarrow} Z = Z_0[A = 0]$. This instructs us to set:
\begin{equation}
    \mathcal{N}_{\mathcal{R}} = \sqrt{\frac{1 + \mu \lambda C_2/2N}{\mu \pi}} \,.
\end{equation}
Then the $T\bar{T}$ deformed partition function for YM$_2$ living on arbitrary manifold is:
\begin{equation}\label{mainresult}
    Z = \sum_\mathcal{R} (\text{dim} \, \mathcal{R})^\chi \, e^{- \frac{\lambda A}{2N} \left(\frac{C_2}{1 + \mu \lambda C_2 / 2 N} \right)} \,.
\end{equation}

This result is in agreement with existing literature. The classical analysis of \cite{Conti2018} found that the Hamiltonian density of $T\bar{T}$ deformed YM$_2$ is given by:
\begin{equation}
    \mathcal{H} = \frac{\mathcal{H}_0}{1 + \mu \mathcal{H}_0} = \frac{\lambda C_2/2N}{1 + \mu \lambda C_2/2N} \,,
\end{equation}
from which they argued that the effect of the $T\bar{T}$ deformation should be incorporated via a redefinition of the quadratic Casimir eigenvalues:
\begin{equation}
    C_2(\mathcal{R}) \rightarrow \frac{C_2(\mathcal{R})}{1 + \mu \lambda C_2 (\mathcal{R})/2N} \,.
\end{equation}
Replacing the Casimir eigenvalues in Eq.\,(\ref{Z0}) with the dressed versions indeed yields Eq.\,(\ref{mainresult}). 

Note that the $T\bar{T}$ deformation of YM$_{2}$ is invariant under area preserving diffeomorphisms, just like the undeformed theory. This fact will pay dividends in our computation of the entanglement entropy in section \ref{ees}.

As another consistency check, the $T\bar{T}$ deformed partition function Eq.\,(\ref{mainresult}) can be shown to satisfy the flow equation:
\begin{equation}
    \partial_\mu Z = A \partial_A^2 Z \,,
\end{equation}
first derived by Cardy in \cite{Cardy2018} for the special case of flat background geometry. To see why our deformed partition function on arbitrary manifold satisfies the flat-space Cardy result, we now turn to a derivation of the flow equation for generalised background geometry. We will find that the presence of contact terms in this generalised flow equation, reflecting operator ordering ambiguities, has an intimate connection with the choice of normalization in the integral transform method. 

\subsection{Flow Equation}\label{flowequation}
As mentioned before, the method used in most studies of the $T\bar{T}$ deformation of quantum field theories is to obtain quantities such as the partition function from solving a flow equation. In this section, we will see how this is in fact equivalent to the method described above by deriving the flow equation that the deformed partition function defined in Eq.\,(\ref{deformation}) satisfies. 

First, a note of caution: the flow equation we describe here should not be seen as a renormalization group flow equation. Instead, it is an equation describing the response of the partition function to tuning the coupling of one irrelevant operator in the theory. The full renormalization group flow equation, or Callan--Symanzik equation, would encode the dependence of the theory on all scales that are present. 
In our case of interest, the undeformed partition function depends only on the area, $Z_0[A(e)] = Z_0[A]$. By applying the integral transform, we know now that the deformed partition function also depends only on the area, $Z[A(f)] = Z[A]$. We now take our final result Eq.\,(\ref{mainresult}), and ask what flow equation it satisfies. We find this to be:
\begin{equation}
\partial_{\mu}Z=A\partial^{2}_{A}Z.
\end{equation}

Note that the form of this equation was sensitive to our choice of the normalization constants $\mathcal{N}_\mathcal{R}$. It is also sensitive to the fact that we choose them separately for each integral. For instance, if we had chosen the same normalization constant for every Gaussian integral--say $\mathcal{N}=1/\sqrt{\pi \mu}$, for example--then we would have obtained the partition function: 
\begin{equation}
Z'=\sum_{\mathcal{R}} \frac{(\textrm{dim} \mathcal{R})^{\chi}}{\sqrt{1+\frac{\mu \lambda C_{2}}{2N}}} e^{-\frac{\lambda A}{2N}\left(\frac{C_{2}}{1+\mu \lambda C_{2}/2N}\right)}\,,
\end{equation}
which satisfies the flow equation:
\begin{equation}
\partial_{\mu}Z'=A\partial^{2}_{A}Z'+\frac{1}{2}\partial_{A}Z'. 
\end{equation}

The extra first derivative term on the right hand side comes form an alternative ordering prescription for the second area derivative term. In particular, 
\begin{equation}
A\partial^{2}_{A}Z'+\frac{1}{2}\partial_{A}Z'=\sqrt{A}\partial_{A}(\sqrt{A}\partial_{A}Z'). 
\end{equation}
This is entirely analogous to the problem of finding a quantization for a phase space function of the form $xp^{2}$ in some mechanical system.

In all, we see that the ordering ambiguity in the flow equation language is intimately tied to the choice of normalization when using the integral transformation \eqref{deformation} to define the deformed partition function.

\section{Entanglement Entropy and the String Expansion}\label{e&s}

Our result for the deformed partition function on a general background gives us a starting point to explore a number of YM$_2$ related phenomena under the $T\bar{T}$ deformation. We begin by computing entanglement entropy for the theory in a general state. Then, specialising to the case of the theory on a sphere, we make contact with existing calculations of entanglement entropy for the Hartle--Hawking state of the undeformed theory in the literature. YM$_2$ also admits a string theoretic description in the large $N$ limit, which is particularly well understood for the case that the background manifold is a sphere. We can then ask whether some semblance of this picture survives in the deformed theory. The sphere partition function will serve as a starting point for this analysis. 

\subsection{Entanglement Entropy}\label{ees}

By making use of the replica trick, we can obtain the entanglement entropy in deformed 2-dimensional Yang--Mills theory directly from Eq.\,(\ref{mainresult}). Further, due to the simple dependence on the area and Euler characteristic of the background manifold, we can actually obtain a general expression valid for any state of the theory which can be prepared via Euclidean path integral \cite{willnicosyd}. Only at the end will we specialise to the Hartle--Hawking vacuum state, for which some results are already known in the case of undeformed YM$_2$. 

To set the stage for the entanglement entropy calculation, we will take our Yang-Mills theory to live on the 2-dimensional background manifold $\mathcal{M}$, and imagine preparing the state $\ket{\Psi}$ by performing a Euclidean path integral over $\mathcal{M}$. Now, we spatially partition the system into a region of interest $\mathcal{A}$ and its complement $\bar{\mathcal{A}}$. Curiously, because the $T\bar{T}$ deformation of YM$_2$ is still invariant under area preserving diffeomorphisms, it does not actually matter here how we choose to partition our system. To $\mathcal{A}$, we can associate a reduced density matrix $\rho_\mathcal{A} = \Tr_{\bar{\mathcal{A}}} \dyad{\Psi}$ obtained by tracing over the degrees of freedom in $\bar{\mathcal{A}}$. Then we can quantify the amount of entanglement between $\mathcal{A}$ and the rest of the system via an application of the von Neumann entropy formula. This gives the entanglement entropy:
\begin{equation}
    S_\mathcal{A} = - \Tr_\mathcal{A} \rho_\mathcal{A} \log \rho_{\mathcal{A}} \,.\label{entropy}
\end{equation}

In practice, computing $S_\mathcal{A}$ directly from Eq.\,(\ref{entropy}) is impossible in all but the most trivial cases, since it involves taking the logarithm of an operator in a potentially infinite system. Instead, we will arrive at $S_\mathcal{A}$ in a slightly more circuitous manner. The key observation is that Eq.\,(\ref{entropy}) can be written equivalently as:
\begin{equation}\label{entropy2}
    S_\mathcal{A} = - \partial_n \left( \Tr \rho_A^n \right) \, \rvert_{n = 1} \,.
\end{equation}
At first, this might not seem to be a simplification; after all, now we're tasked with computing moments of the reduced density matrix. Luckily, though, the replica trick gives us a way to do just that (see \cite{HEEbook} for a review). More precisely, it gives us a way to relate the trace of $\rho_\mathcal{A}^n$ to the partition function on a replicated manifold. 

The replica trick instructs us to first take our spacetime of interest, the background manifold $\mathcal{M}$, and to replicate it such that there are $n$ total copies. For each copy, we partition into subsystems $\mathcal{A}$ and $\mathcal{\bar{A}}$ by making a cut along $\mathcal{A}$. We will denote the upper boundary of the cut on the $i^{th}$ copy as $\mathcal{A}^+_i$ and the lower boundary as $\mathcal{A}^-_i$. The replicated manifold $\mathcal{M}_n$ can then be constructed by ``glueing" the $n$-sheets together cyclically along the cuts. That is, we identify $\mathcal{A}^+_{i-1} \leftrightarrow \mathcal{A}^-_{i}$, $\mathcal{A}^+_{i} \leftrightarrow \mathcal{A}^-_{i+1}$, \textit{etc}. The resultant manifold $\mathcal{M}_n$ is the $n$-fold branched cover over $\mathcal{M}$.

On each copy of the background $\mathcal{M}$, we compute the reduced density matrix $\rho_\mathcal{A}$ via (Euclidean) path integral. The glueing operation by which we construct $\mathcal{M}_n$ is then morally equivalent to taking the trace over the $n$ copies of $\rho_\mathcal{A}$. This can equivalently be seen as computing the partition function $Z_n$ on the $n$-fold cover $\mathcal{M}_n$ (up to normalisation). The formal result is that:
\begin{equation}
    \Tr \rho_\mathcal{A}^n = \frac{Z_n}{(Z_1)^n} \,,
\end{equation}
where $Z_1$ is the partition function of the original theory on $\mathcal{M}$. 

The one missing ingredient is the deformed partition function on the $n$-fold cover, $Z_n$. This is easy enough to obtain since Eq.\,(\ref{mainresult}) depends only on the total area and Euler characteristic. On $\mathcal{M}_n$, these quantities are given by \cite{Donnelly:2014}:
\begin{subequations}
\begin{equation}
    A_n = A n \,,
\end{equation}
\begin{equation}
    \chi_n = 2 n + 2(1-n) m \,,
\end{equation}
\end{subequations}
where $A$ is the total area of a single copy of the replicated manifold and $m$ is the number of partitions into which we've subdivided our system. Here, $m = 1$, so we simply have $\chi_n = 2$. Finally, before applying Eq.\,(\ref{entropy2}), let us write $Z_n$ in a slightly more suggestive form by introducing the (normalised) probability distribution for the representations:
\begin{equation}\label{probability}
    P(\mathcal{R}) = \frac{1}{Z_1} (\text{dim} \mathcal{R})^\chi e^{- \frac{\lambda C_2/2N}{1 + \mu \lambda C_2/2N} A} \,.
\end{equation}
In terms of $P(\mathcal{R})$, we have:
\begin{equation}
    Z_n = \sum_\mathcal{R} (\text{dim} \mathcal{R})^2 e^{- \frac{\lambda C_2/2N}{1 + \mu \lambda C_2/2N} A n} = \sum_\mathcal{R} (\text{dim} \mathcal{R})^{2 - n \chi} (Z_1)^n P(\mathcal{R})^n \,.
\end{equation}
Then applying Eq.\,(\ref{entropy2}) gives:
\begin{equation}
\begin{split}
    S_\mathcal{A} & = - \partial_n \left( \frac{Z_n}{(Z_1)^n} \right) \bigg\rvert_{n = 1}\\
    & = - \partial_n \left( \sum_\mathcal{R} (\text{dim} \mathcal{R})^{2 - n \chi} P(\mathcal{R})^n \right) \bigg\rvert_{n = 1}\\
    & = \sum_\mathcal{R} P(\mathcal{R}) (\text{dim} \mathcal{R})^{2-\chi} \left[ \chi \log (\text{dim} \mathcal{R}) - \log P(\mathcal{R}) \right] \,.
\end{split}
\end{equation}

Again, because YM$_2$ is semi-topological, the actual bipartition into $\mathcal{A}$ and $\bar{\mathcal{A}}$ is immaterial. We can then say that the entanglement entropy of $T\bar{T}$ deformed 2-dimensional Yang--Mills in a completely arbitrary state $\ket{\Psi}$ prepared via Euclidean path integral is:
\begin{equation}\label{arbitraryee}
    S = \sum_\mathcal{R} P(\mathcal{R}) (\text{dim} \mathcal{R})^{2-\chi} \left[ \chi \log (\text{dim} \mathcal{R}) - \log P(\mathcal{R}) \right] \,,
\end{equation}
with the probability distribution of representations $P(\mathcal{R})$ given by Eq.\,(\ref{probability}). 

Now to make contact with existing results in the literature, let us restrict to the case that the background manifold is the two-sphere $\mathbb{S}^2$, for which $\chi = 2$. Slicing the sphere in angular time and taking the path integral over this hemisphere geometry gives the Hartle--Hawking vacuum state $\ket{HH}$. The corresponding entanglement entropy for $T\bar{T}$ deformed YM$_2$ in this state is:
\begin{equation}\label{HHee}
    S_{HH} = \sum_\mathcal{R} \left[ 2 P(\mathcal{R}) \log (\text{dim} \mathcal{R}) - P(\mathcal{R}) \log P(\mathcal{R}) \right] \,,
\end{equation}
where again, $P(\mathcal{R})$ is given by Eq.\,(\ref{probability}) with $\chi = 2$. Meanwhile, for the undeformed theory, entanglement entropy in the Hartle--Hawking state is given by \cite{Donnelly:2014} \cite{GROMOV201460}:
\begin{equation}\label{HHee0}
    (S_0)_{HH} = \sum_\mathcal{R} \left[ 2 P_0(\mathcal{R}) \log (\text{dim} \mathcal{R}) - P_0(\mathcal{R}) \log P_0(\mathcal{R}) \right] \,,
\end{equation}
where now 
\begin{equation}
    P_0 = \frac{1}{(Z_0)_1} (\text{dim} \mathcal{R})^2 e^{- \frac{\lambda C_2}{2N} A} \,,
\end{equation}
is the probability distribution of representations in the undeformed theory. We see that Eqs.\,(\ref{HHee}) and (\ref{HHee0}) take the same structural form. To understand why this is, we need to first understand the origin of the terms in these formulae. 

In both expressions for $S_{HH}$, the second term, $-\sum_\mathcal{R} P(\mathcal{R}) \log P(\mathcal{R})$, is essentially the classical entropy. The only observables one can measure in YM$_2$ are gauge invariant functions of the (non-Abelian) electric field $E^a(x)$. The Gauss law constraint sets $E^a(x)$ to be constant over the whole circle, so measurements of gauge invariant observables constructed out of $E^a$ made in region $\mathcal{A}$ are always going to be correlated with those made in $\bar{\mathcal{A}}$. This is another way to understand why the manner in which we partitioned the circle into $\mathcal{A}$ and $\bar{\mathcal{A}}$ didn't matter. Since the measurements are always correlated, tracing over $\bar{\mathcal{A}}$ should introduce no additional uncertainty. One would then expect that the only contribution to the entropy would be the statistical entropy coming from the classical uncertainty in the outcome of measurements of electric field observables. 

There is another source of entropy, though, as evidenced by the first term in Eqs.\,(\ref{HHee}) and (\ref{HHee0}). This term comes from counting ``edge modes"--additional degrees of freedom coming from states at the endpoints which transform non-trivially under gauge. The question of whether such non-gauge invariant degrees of freedom should be included in the definition of entanglement entropy is discussed at length in \cite{Donnelly:2014}, with an argument made in the affirmative. Here, we adopt this stance as well, and so each of the two endpoints of region $\mathcal{A}$ is taken to contribute a factor $\log (\text{dim} \mathcal{R})$, leading to the $2 \sum_\mathcal{R} P(\mathcal{R}) \log (\text{dim} \mathcal{R})$ term in entanglement entropy. 


Note that aside from these edge mode terms, there are no terms in the entanglement entropy coming from local degrees of freedom. In fact, even in the original undeformed theory, entanglement entropy was already finite. This lack of UV divergences makes sense, given that YM$_2$ has no local propagating degrees of freedom\footnote{The existence of local gauge invariant degrees of freedom is obstructed by YM$_2$'s extended symmetry group of area-preserving diffeomorphisms. The theory can still have degrees of freedom; to see them, though, one needs to look at spacetimes with non-trivial topology and non-local quantities like Wilson loops \cite{Cordes:1994}.}. In this sense, the effect of $T\bar{T}$ on entanglement entropy is somewhat obscured by the triviality of our theory. The deformation has been argued to act as an effective UV cutoff, rendering quantities like entanglement entropy finite. In the context of the $T\bar{T}$ deformation of large $c$ conformal field theories, this was shown to be the case in \cite{Donnelly:2018bef}. This result was further generalized in   \cite{Lewkowycz:2019xse} and a similar effect is seen in higher dimensional generalizations \cite{Grieninger:2019zts}. For YM$_2$, which is already UV finite, the effect of the deformation on entanglement entropy is then minimal. Eqs.\,(\ref{HHee}) and (\ref{HHee0}) maintain the same structural form, including terms coming from classical statistical entropy as well as entropy associated with counting edge modes. The only difference is that the probability distributions associated with the irreducible representations $\mathcal{R}$ are shifted due to the deformation's dressing of the Casimir eigenvalues.

\subsection{Fate of the string picture}

One aspect of 2-dimensional Yang--Mills which makes it an exceptionally interesting theory in spite of its lack of local degrees of freedom and semi-topological nature is the fact that it admits a dual string interpretation. As first shown by Gross and Taylor \cite{Gross:1993}, YM$_2$ \textit{is} a string theory in the large $N$ limit. More precisely, expanding the partition function in a power series in $1/N$ results in an expression whose coefficients match order-by-order with those expected from a sum over maps from a 2D covering space $\Sigma$ to a 2D target space $\mathcal{M}$, each map being weighted by the Nambu-Goto action. Provided the identification of $1/N$ with the string coupling $g_s$ and $\lambda$ with the string tension $1/2 \pi \alpha'$, we have the moral equivalence:
\begin{equation}
    \log Z_{YM} [N, \lambda, A] = Z_{string}\left[ g_s = \frac{1}{N}, \alpha' = \frac{1}{\pi \lambda} \right]\,.
\end{equation}

The intuitive picture is that we have a closed string worldsheet $\Sigma$ wrapping $n$ times around the manifold $\mathcal{M}$. Branch points on $\Sigma$ correspond to interactions where strings can either split apart or join together. In addition to these elementary branch point singularities, manifolds with non-vanishing $\chi$ also admit so-called $\Omega$ and $\Omega^{-1}$ point singularities, the number of which is fixed by the Euler characteristic of $\mathcal{M}$. Only recently have these singularities been given an interpretation\footnote{The authors of \cite{Donnelly:2019} have argued that $\Omega$ and $\Omega^{-1}$ points are related to positive and negative index singularities, respectively, of the modular flow.} in the context of the string picture \cite{Donnelly:2019}. In the case of the two-sphere $\mathcal{M} = \mathbb{S}^2$, for which $\chi = 2$ and there are two $\Omega$ points, the interpretation is particularly nice. This is the case to which we'll restrict our analysis.

For a single chiral sector\footnote{The closed string Hilbert space $\mathcal{H}$ is a subspace of the tensor product of chiral and anti-chiral sectors $\mathcal{H}^+ \otimes \mathcal{H}^-$. These correspond to strings winding in opposite directions.} of YM$_2$ living on $\mathbb{S}^2$, the partition function can be expanded in string basis as \cite{Donnelly2017}:
\begin{equation}\label{undeformedstrings}
    Z_{YM} = \sum_n \frac{1}{n!} \sum_k \frac{1}{k!} \left( - \frac{n \lambda A}{2} \right)^k \sum_r \frac{(-1)^r}{r!} \left( \frac{\lambda A}{N} \right)^r \sum_{\sigma \in S_n} \sum_{p_1...p_r \in T_2} N^{K_\sigma} N^{K_{p_1...p_r \sigma}} \,.
\end{equation}
As promised, we are summing over an $n$-sheeted covering, with the $1/n!$ accounting for redundancy in summing over homomorphisms differing only by a trivial relabeling. The $\sum_k \frac{1}{k!} \left( - \frac{n \lambda A}{2} \right)^k = \exp (- n \lambda A/2)$ is the Nambu-Goto action of the string worldsheet wrapping $n$ times about the sphere and describes the ``free" part of the theory. Specifically, $nA$ is the area of the string worldsheet with no foldings, and $\lambda$ is proportional to the string tension. The $r$ string interactions are encoded in $\sum_r \frac{(-1)^r}{r!} \left( \frac{\lambda A}{N} \right)^r$. Here, the $1/r!$ accounts for the indistinguishability of the interactions, a factor of the string coupling $1/N = g_s$ accompanies each interaction, and $\lambda A$ is a ``modulus factor" obtained from integrating over all possible places where an interaction could take place\footnote{The $(-1)^r$ factor is thought to relate to the fermionic nature of interaction points \cite{Donnelly2017}, but a precise identification remains unclear.}. Finally, $N^{K_\sigma}$ accounts for the $K_\sigma$ closed strings in the initial state $\ket{\sigma}$ emitted from one $\Omega$ point, while $N^{K_{p_1...p_r \sigma}}$ accounts for the $K_{p_1...p_r \sigma}$ closed strings in the final state $\ket{p_1...p_r \sigma}$ absorbed at the other $\Omega$ point. We sum over all permutations $\sigma \in S_n$ for the initial state as well as all sequences of transpositions $p_1...p_r \in T_2$ leading to the final state. 

The overall picture, then, is that we have the following evolution in Euclidean time: An initial state $\ket{\sigma}$ of $K_\sigma$ closed strings is emitted from one $\Omega$ point. The strings undergo $r$ interactions which locally cut and re-glue them, acting as a series of transpositions $p_1 p_2...p_r \equiv p \in T_2$ which take $\ket{\sigma} \rightarrow \ket{p \sigma}$. The interactions may change individual winding numbers, but preserve total winding number. The resultant final state $\ket{p \sigma}$ of $K_{p \sigma}$ strings is absorbed at the other $\Omega$ point. 

A natural question would be what happens to this string picture upon deforming the theory with $T\bar{T}$. After all, it has been shown that deforming a theory of free massless bosons with $T\bar{T}$ results in the Nambu-Goto action in static gauge \cite{Cavaglia:2016oda}. In this result as well as in other contexts (\textit{e.g.}, \cite{Kraus:2018xrn}, \cite{Dubovsky:2018bmo}, \cite{Giveon:2017nie}), it has become apparent that deforming QFTs with $T\bar{T}$ results in theories which are in a sense non-local. It would then be interesting to see what happens starting from a theory of strings.

There are actually two convenient ways to go about obtaining the string expansion of the $T\bar{T}$ deformed partition function. One would be to start from the undeformed partition function in string basis and apply the kernel integral transform. While computationally simple, the resultant combinatorics make interpreting the expression rather difficult. On the other hand, given that we can identify the form of the deformed Hamiltonian density based on the transformation of the quadratic Casimirs under the deformation,
\begin{equation}
    \mathcal{H}_0 \rightarrow \mathcal{H} = \frac{\mathcal{H}_0}{1 + \mu \mathcal{H}_0} = \frac{(\lambda/2N) \hat{C}_2}{1 + (\mu \lambda/2N) \hat{C}_2} \,,
\end{equation}
we could also obtain $Z$ as the Euclidean evolution between initial and final states inserted at the $\Omega$ points:
\begin{equation}\label{evolution}
\begin{split}
    Z & = \matrixel{\Omega}{e^{-\beta \mathcal{H}}}{\Omega} \\
    & = \matrixel**{\Omega}{e^{- \frac{(\lambda A/2N)\hat{C}_2}{1+ (\mu \lambda/2N)\hat{C}_2}}}{\Omega} \\
    & = \sum_k \frac{(-1)^k}{k!} \left(\frac{\lambda A}{2N}\right)^k \sum_\ell \frac{(-1)^\ell}{\ell!} \left( \frac{\mu \lambda}{2N} \right)^\ell \frac{(k + \ell - 1)!}{(k - 1)!} \matrixel{\Omega}{\hat{C}_2^{k+\ell}}{\Omega} \,.
\end{split}
\end{equation}

In order to evaluate the matrix element, we note that the state $\ket{\Omega}$ may be written in string basis\footnote{In the representation basis, $\ket{\Omega} = \sum_\mathcal{R} \text{dim} \, \mathcal{R} \ket{\mathcal{R}}$. The bases are related by the Frobenius relation $\ket{\mathcal{R}} = \sum_{\sigma \in S_n} \chi_\mathcal{R}(\sigma)/n! \ket{\sigma}$, where $\chi_\mathcal{R}(\sigma)$ is the character of the permutation group associated to representation $\mathcal{R}$. See \cite{Donnelly2017} for more technical details.} as:
\begin{equation}
    \ket{\Omega} = \sum_n \frac{1}{n!} \sum_{\sigma \in S_n} N^{K_\sigma} \ket{\sigma} \,,
\end{equation}
with $\sigma$ a permutation in the symmetric group $S_n$ and $K_\sigma$ the number of closed strings in the initial state $\ket{\sigma}$. Meanwhile, the quadratic Casimir operator can be decomposed in terms of ``free" and ``interacting" parts: $\hat{C}_2 = Nn + 2\hat{C}_{int}$. The leading term counts the total winding number $n$ while the interaction term implements a transposition $p \in T_2$, with the factor of 2 accounting for double counting. That is,
\begin{equation}
    \hat{C}_2 \ket{\sigma} = Nn\ket{\sigma} + 2 \sum_{p \in T_2} \ket{p \sigma} \,.
\end{equation}
Then, making use of the fact that the inner product is $\braket{\Omega}{\sigma} = N^{K_\sigma}$, we have:
\begin{equation}
    \begin{split}
        \matrixel{\Omega}{\hat{C}_2}{\Omega} & = \sum_n \frac{1}{n!} \sum_{\sigma \in S_n} N^{K_\sigma} \matrixel{\Omega}{\hat{C}_2^{k+\ell}}{\sigma}\\
        & = \sum_n \frac{1}{n!} \sum_{\sigma \in S_n} N^{K_\sigma} \sum_{r = 0}^k \frac{k!}{r! (k-r)!} (Nn)^{k-r} \sum_{s = 0}^\ell \frac{\ell!}{s! (\ell - s)!} (Nn)^{\ell - s} 2^{r+s} \matrixel{\Omega}{\hat{C}^{r+s}_{int}}{\sigma} \\
        & = \sum_n \frac{1}{n!} \sum_{\sigma \in S_n} N^{K_\sigma} \sum_{r = 0}^k \frac{k!}{r! (k-r)!} (Nn)^{k-r} \sum_{s = 0}^\ell \frac{\ell!}{s! (\ell - s)!} (Nn)^{\ell - s} 2^{r+s} \sum_{p_1...p_{r+s} \in T_2} N^{K_{p_1...p_{r+s} \sigma}} \,.
    \end{split}
\end{equation}
Plugging into Eq.\,(\ref{evolution}) and regrouping terms, we arrive at the following expression for the string basis expansion of the $T\bar{T}$ deformed partition function:
\begin{equation}\label{deformedstrings}
\begin{split}
    Z = \sum_n \frac{1}{n!} \sum_k \frac{1}{k!} \left( - \frac{n \lambda A}{2} \right)^k \sum_r & \frac{(-1)^r}{r!} \left( \frac{\lambda A}{N} \right)^r \sum_\ell \frac{1}{\ell !} \left( - \frac{n \lambda \mu}{2} \right)^\ell \sum_s \frac{(-1)^s}{s!} \left( \frac{\lambda \mu}{N} \right)^s \\
    & \frac{(k + r + \ell + s - 1)!}{(k + r - 1)!} \sum_{\sigma \in S_n} \sum_{p_1...p_{r+s} \in T_2} N^{K_\sigma} N^{K_{p_1...p_{r+s} \sigma}} \,.
\end{split}
\end{equation}

The overall form of Eq.\,(\ref{deformedstrings}) bears a structural resemblance to its undeformed counterpart Eq.\,(\ref{undeformedstrings}). Nevertheless, the interpretation of this expression is problematic. The $1/N$ expansion can still be interpreted as a sum over maps from an $n$-sheeted covering space $\Sigma$ to a target space $\mathcal{M}$, however the weighting of these maps is a lot more complicated. In the undeformed theory, each map was simply weighted by the (free) Nambu-Goto action. The remaining terms in the expansion were then interpreted as string interactions. After deforming the theory with $T\bar{T}$, though, there is a non-trivial coupling between the ``free" and ``interacting" sectors. Unfortunately, this precludes a straightforward interpretation. It somewhat looks like at each interaction site indexed by $r$, at which strings can split and rejoin, there is the possibility for a second kind of sub-interaction, indexed by $s$. This sub-interaction would also allow for strings to split and rejoin, but would occur over an area with scale set by $\mu$. This picture is tenuous, however, and we leave the precise interpretation of this expression and the fate of the string picture to future investigation.

\section{Discussion}
In this article, we have advocated for Eq.\,(\ref{deformation}) as an alternative definition for the action of the $T\bar{T}$ deformation which should apply for any 2-dimensional quantum field theory on general background. 

This integral transform was first put forth in the context of holography in \cite{Freidel:2008sh} as a means of relating the partition function of a 2D conformal field theory with the 3D gravity bulk wave function satisfying the radial Wheeler-DeWitt equation. It was then noted in \cite{McGough:2016lol} that this integral transform looks like the $T\bar{T}$ deformation for CFTs in the sense that the flow equation usually used to define the $T\bar{T}$ deformation of CFTs takes the form of the WDW equation in three dimensions with a negative cosmological constant. These authors thus argued that via this transform, one can obtain the $T\bar{T}$ deformed QFT partition function $Z_{QFT}[f]$ from that of an undeformed CFT $Z_{CFT}[e]$. 

Motivated by the observation that taking the $\mu$ derivative of $Z[f]$ as defined by Eq.\,(\ref{deformation}) effectively pulls down the $T\bar{T}$ operator, we have proposed that this integral transform be extended as a definition of the deformation for general 2D QFTs. That it involves an integral over frame fields is reminiscent of the proposal in \cite{Dubovsky:2018bmo}, which was used to obtain the torus partition function. The precise connection between the two approaches is dealt with in more detail in \cite{mazenc2019t}. We have supported our proposal by applying the integral transform in the test case of 2-dimensional Yang--Mills theory, whose partition function has a very simple Gaussian dependence on the frame fields. The resultant $T\bar{T}$ deformed partition function matches that obtained via other methods for all cases where the result was previously known, but can now be extended to any background manifold. 

Implicit in the kernel definition is an ambiguity regarding normalisation. For the test case of YM$_2$ investigated in this article, we chose a normalisation based on physical arguments that the deformation should do nothing to the theory in the topological limit, for which the stress-tensor vanishes. Consequently, the deformed partition function obeyed the standard Cardy flow equation \cite{Cardy2018} with no first derivative term. However, in formulating a more general flow equation for the theory on arbitrary background, we found that such terms could arise depending on the order in which functional derivatives were taken at coincident points. This ordering ambiguity in the flow equation formalism translates to the normalisation ambiguity in the integral transform definition of the deformation. Further development of the kernel method should hopefully make this connection more precise. 

From Eq.\,(\ref{mainresult}), the entanglement entropy of deformed YM$_2$ in any arbitrary state could be computed. Specialising to the case of the Hartle--Hawking vacuum state, the entanglement entropy Eq.\,(\ref{HHee}) was found to take the same structural form as in the undeformed theory, Eq.\,(\ref{HHee0}). In both, there was a term arising from classical statistical entropy as well as a term coming from the counting of edge modes. The only difference lay in the probability distributions associated with the irreducible representations, which were shifted due to the deformation's dressing of the quadratic Casimir eigenvalues. This was in a sense to be expected. The $T\bar{T}$ deformation has been argued to act as an effective UV cutoff on entanglement entropy \cite{Donnelly:2018bef}. However, YM$_2$, being a semi-topological theory with no local degrees of freedom, is already UV finite. Thus, it is reasonable that the deformation should have a minimal effect on entanglement entropy. 

Because the deformed Yang--Mills partition function has such a simple dependence on the geometry, there are unfortunately not many observables which can be computed from it. In this article, we have calculated the entanglement entropy of the deformed theory in an arbitrary state, but it would be interesting to see what other observables one can in principle compute. In the undeformed theory, another set of interesting observables considered are correlation functions of Wilson loops. Perhaps such observables can be computed in the deformed theory as well.
Despite its simplicity, Yang--Mills does have a number of interesting associated phenomena which can be re-examined under the prism of $T\bar{T}$. One of these, the third order Douglas--Kazakov phase transition, was investigated in \cite{Santilli2019}. By performing the large $N$ analysis, it was found that not only does the transition persist for a range of deformation parameter values, but the interpretation of the transition as being induced by unstable instantons remains valid. Another interesting aspect of large $N$ Yang--Mills which had not yet been examined in the deformed theory was its dual interpretation as a string theory. 

Upon performing the large $N$ string expansion of the deformed partition function, we arrived at Eq.\,(\ref{deformedstrings}). While this expression superficially looks quite similar to Eq.\,(\ref{undeformedstrings}), its interpretation is troubled by the factorial factor which acts to couple the ``free" and ``interacting" terms of the theory. It is possible that this is simply an exotic string theory with a very complex weighting. If we take this perspective, then it looks like we can have two types of interactions which, while both splitting and rejoining strings, occur over two different fundamental length scales. Or it could well be that $T\bar{T}$ deformed Yang--Mills is simply no longer a string theory.
We invite the reader to investigate this question further.

\section*{Acknowledgements}
We would like to thank W. Donnelly, N. Vald\'{e}s, S. Timmerman, E. Mazenc and R. Soni for many fruitful discussions. We would also like to thank A. Tolley for clarifications regarding the choice of the path integral measure. 
This research was supported in part by Perimeter Institute for Theoretical Physics. Research at Perimeter Institute is supported in part by the Government of Canada through the Department of Innovation, Science and Economic Development Canada and by the Province of Ontario through the Ministry of Economic Development, Job Creation and Trade.

\bibliographystyle{utphys}
\bibliography{ttbar2dym}

\end{document}